# Point Defects and Their Dynamic Behaviors in Silver Monolayer Intercalated between Graphene and SiC

Van Dong Pham,[1*] Arpit Jain[2], Chengye Dong[2,3,4], Li-Syuan Lu[2], Joshua A. Robinson[2,3,4,5], Achim Trampert[1], Roman Engel-Herbert[1]

[1]*Paul-Drude-Institut für Festkörperelektronik, Leibniz-Institut im Forschungsverbund Berlin e. V., Hausvogteiplatz 5-7, 10117 Berlin, Germany*

[2]*Department of Materials Science and Engineering, The Pennsylvania State University, University Park, Pennsylvania 16802, United States*

[3]*2-Dimensional Crystal Consortium, The Pennsylvania State University, University Park, Pennsylvania 16802, United States*

[4]*Center for 2-Dimensional and Layered Materials, The Pennsylvania State University, University Park, Pennsylvania 16802, United States*

[5]*Center for Atomically Thin Multifunctional Coatings, The Pennsylvania State University, University Park, Pennsylvania 16802, United States*

**ABSTRACT:** Point defects give rise to sharp modifications in the structures and electronic properties of two-dimensional metals, offering an atomic-level platform for fundamental studies and potential applications. In this work, we investigate atomic-scale defects in a two-dimensional silver monolayer intercalated between epitaxial graphene and SiC using scanning tunneling microscopy. Dark and bright defects are identified as vacancies or substitutional impurities within the silver monolayer, each hosting a localized electronic state. Remarkably, under tunneling electron excitation at negative bias, the bright defects exhibit dynamic behaviors characterized by inelastic switching between two states. The switching can be reversibly controlled by the microscope tip, enabling the defects to function as atomic-scale two-level conductance switches. Analysis of defect switching reveals possible defect origins and the relationship between dark and bright defect species. Our findings establish a pathway to precise manipulation of defects in two-dimensional metals and uncover previously unexplored dynamics with potential use in nanoelectronics.

[*]Email: pham@pdi-berlin.de

Point defects in solid crystals have attracted extensive attention in various fields because they not only alter the host material properties but can give rise to striking quantum behaviors, offering new platforms for fundamental studies and quantum technologies[1]. In bulk crystals, nitrogen-vacancy color centers in diamond[2] and in SiC[3] can act as quantum light sources. Defects in two-dimensional (2D) materials, likewise, exhibit distinct phenomena. Vacancy defects in graphene, for instance, strongly reduce charge-carrier mobility and induce local magnetic moments[4]. Defects can act as single-photon emitters in WSe2[5], exhibit an anomalous Zeeman shift under a magnetic field in antiferromagnetic FeSn films[6] or mediate new magnetic ordering in 2D transition metal phosphorus trisulfides[7], to mention a few.

Atomically thin metals intercalated at the epitaxial graphene/SiC(0001) interface have emerged as a promising class of 2D materials with strong potential for quantum and nanoelectronic applications, owing to their wafer-scale integration and environmental stability[8–10]. As in other 2D materials, point defects are ubiquitous in intercalated metals and strongly modify their structural and electronic properties[11–13]. However, the atomic-scale structures, electronic, and dynamic properties of these defects remain mostly unexplored. Gaining such understanding could enable atomic-level control over their behaviors in atomically confined metal at the graphene/SiC interface, and establish a foundation for defect engineering in 2D metal systems.

Scanning tunneling microscope (STM) offers a powerful tool for this purpose, allowing identification and controlled manipulation of single atoms on a surface[14,15], hydrogen switching in organic molecules[16–19], and atomic defects[20–23]. These capabilities enable direct identification of the chemical nature, atomic structures and control of defects in intercalated metals, which remain elusive[24,25].

In this Letter, we demonstrate a first experimental attempt to explore the properties and the dynamic behaviors of point defects in a silver (Ag) monolayer intercalated at the graphene/SiC interface using cryogenic STM. Understanding these defects is of particular importance, as confined Ag layers have been found to exhibit intriguing quantum effects[10,26,27]. The defects show a pronounced topographical contrast compared to the defect-free Ag and display localized defect-induced electronic states. Moreover, the STM tip-sample junction serves as a unique experimental setup for controlling electron transport through individual defects. This process induces switching transitions at a defect site, characterized by a two-level tunneling conductance, that can be tuned by tunneling current under a negative bias. These findings provide new insight into defect dynamics in intercalated metals at the graphene/SiC interface, in which the controlled tunneling-driven switching of these defects highlights their potential for quantum technology, including atomic-scale switches[16,17].

Zero-layer graphene (ZLG) was prepared by thermal decomposition of semi-insulating 6H-SiC(0001) (Coherent Corp) via silicon sublimation. The SiC substrate was first annealed at 1400 °C for 30 min in a 10% $H_2$/Ar ambient (700 Torr) to remove surface contamination, followed by annealing at 1600 °C for 30 min in pure Ar (700 Torr) to produce a predominantly ZLG surface with ~10% monolayer graphene overgrowth. Silver intercalation was performed in a quartz-tube furnace (Thermo Scientific Lindberg/Blue M Mini-Mite) by placing high-purity Ag powder (99.99%, Sigma-Aldrich) below the ZLG/SiC sample (facing down) inside an alumina crucible. The system was heated to 900 °C in flowing Ar (500 Torr) for 1 h and then cooled to room temperature, yielding a quasi-freestanding monolayer graphene/Ag/SiC heterostructure. The sample was transported in air and annealed in the STM preparation chamber at 150 °C for 10 min under UHV conditions and transferred into the STM maintained at 5 K.

STM measurements were carried out in a Createc cryogenic STM operating at 5 K. Topography images were acquired in the constant-current mode. Differential conductance (dI/dV) spectra were recorded to probe the local electronic density of states (LDOS) using a lock-in technique with a 5 mV (peak-to-peak) modulation at 675 Hz. Electrochemically etched tungsten tips were used, cleaned by $Ne^+$ ion bombardment, annealed by electron beam heating, and calibrated on an Ag(111) surface.

Intercalated Ag forms as a monolayer with two structural phases under graphene [Supporting Information, S1][28]. Figure 1(a) represents a large-scale STM image (-1.3 V) on one of these phases, where defects exhibit particular dynamic behaviors central to our investigation. At this bias, the graphene overlayer is not visible, whereas the Ag structure is displayed, exhibiting a moiré pattern emerging from the Ag electronic states which overlap with the tip wave function through the graphene cap[29]. The observed moiré pattern exhibits a lattice constant of ~14.7 Å (white rhombus), resulting from the superposition of the Ag monolayer with a lattice constant of 2.98 Å and the graphene overlayer [see more detail in Supporting Information, S2]. In this configuration, each (3√3×3√3)R30°-Ag supercell is in registry with a (5×5)-SiC supercell, yielding a (27:25)-Ag structure, as confirmed by LEED measurements[28].

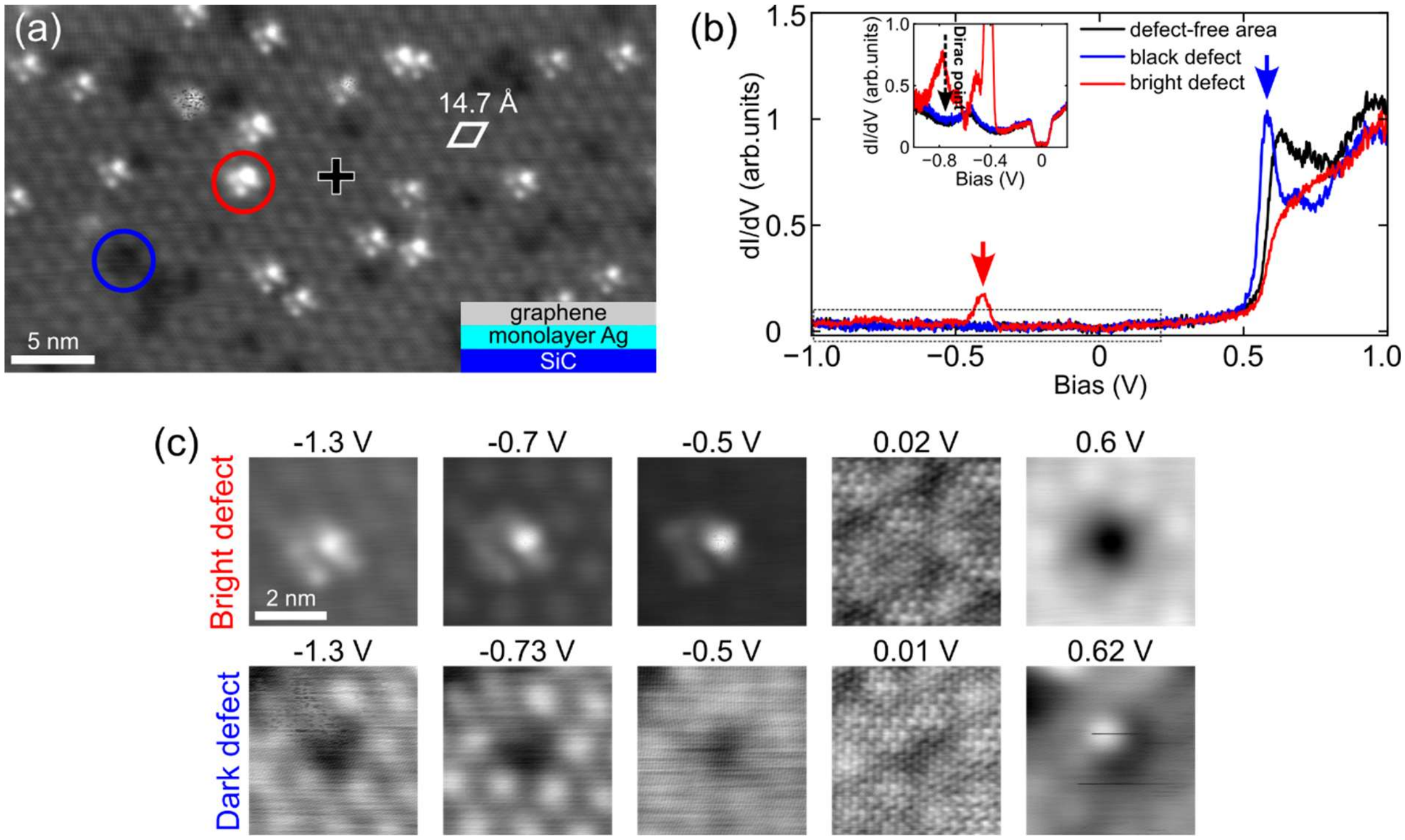

**Figure 1.** (a) Large-scale STM topography image (-1.3 V, 100 pA) of intercalated Ag; the moiré pattern with a periodicity of ~14.7 Å (white rhombus) is an indication of defect-free intercalated Ag below the graphene cap, which results from the superposition of the Ag monolayer with an Ag-Ag spacing of 2.98 Å and the graphene overlayer in which each (3√3×3√3)R30°-Ag supercell is in commensuration with a (5×5)-SiC supercell, denoted as (27:25)-Ag structure. Notably, dark (blue circle) and bright spots (red circle) are identified as point defects within the intercalated Ag monolayer, appearing with distinct contrast. Inset: Schematic illustration of the graphene/Ag/SiC sample. (b) dI/dV spectra acquired above bright (red), black (blue) defects, and defect-free intercalated locations (black), respectively. The occupied peak in the red spectrum taken above the bright defect and the unoccupied peak recorded above the dark defect are indicated

by red and blue arrows, respectively. Inset: Zoomed spectra around the Fermi level (from the dashed rectangle) on these locations, revealing only a dominant LDOS of graphene with a phonon-induced inelastic gap (~134 mV)[30] and a Dirac point at -0.75 eV. (c) Bias-dependent topography STM images of a bright defect (upper panel) and a dark defect (lower panel) acquired with the same tip. The scale bar in the first STM image in (c) is used for all images.

Notably, random bright protrusions (red circle) and dark vacant sites (blue circle) are found coexisting with the moiré pattern at -1.3 V. In contrast, at low sample biases, they are absent, and only the graphene honeycomb lattice is observed (Supporting Information, S3). This observation suggests that the bright and dark species (with a concentration of $\sim 2.2\times10^{12}$ $cm^{-2}$) correspond to point defects in the monolayer Ag which significantly perturbs the Ag lattice, resulting in complete suppression of a bright lobe of the moiré pattern.

Density functional theory (DFT) calculations could elucidate the origin of the observed defects; however, the involvement of multiple defect types would require sophisticated simulations. We therefore focus on the experimental investigation of defect dynamics, with theoretical studies planned for future work. Based on the epitaxial relation between a monoatomic Ag layer and the well-defined SiC, we suggest that the bright and dark defects originate from either Ag vacancies or substitutional Si atoms migrating from the SiC substrate[31–34], or other unknown impurities during the intercalation process.

Differential conductance tunnel spectroscopy (dI/dV), which probes the LDOS, is used to identify the electronic signatures of each defect species. Figure 1(b) shows three dI/dV spectra taken on the bright defect (red spectrum), dark defect (blue), and a defect-free location (black) using the same STM tip. Near the Fermi level (V=0), all spectra (see inset) show a V-shape density of states with the Dirac point identified at -0.75 V[10], associated with a phonon-induced inelastic gap (~134 meV)[30] as displayed in the inset. The preservation of graphene's density of states, even probed at the defect sites, indicates continuous graphene coverage and its weak interaction with the underlying Ag[10].

The electronic signatures of each defect type are reflected well beyond the Fermi level. The spectra taken above the bright and dark defects reveal sharp occupied and unoccupied electronic states, indicated by red and blue arrows, respectively. The positions of these peaks shift slightly, induced by the local environment around each defect. The steep onset at ~0.6 eV probed in a defect-free area (black curve) is attributed to the conduction band minimum of the intercalated Ag[27], which is partially reflected at the defect sites. Since the peaks are solely probed at the defect sites, we attribute them to defect-induced localized states, analogous to the electronic states arising from single vacancies in other 2D materials, such as graphene[35,36], or substitutional oxygen in transition metal dichalcogenide layers[37]. In particular, these peaks do not shift with varying tip height (Supporting Information, S4), ruling out the possibility that they originate from tip-induced charging effect, since charging peaks are known to shift significantly as a function of the tip-sample distance[24,38–40].

Figure 1(c) displays close-up STM topography images of dark and bright defects. At negative bias, each bright defect (upper panel) appears as a protrusion with smaller satellite dots, and they orient in the same direction, indicating identical atomic structure. The satellite spots are likely Ag atoms adjacent to the defect, illuminated under the same imaging condition induced by a local charge transfer with the defect. Despite the triangular arrangement in the Ag layer, the defect may not be located symmetrically at the missing Ag atom due to different atomic sizes, or strain-induced effect, causing an asymmetric configuration as seen in the STM images.

At a positive bias, the same defect appears as a deep vacant site. In contrast, at negative bias, the dark defects (lower panel) appear as empty vacant sites, whereas at positive bias, they appear as atom-like protrusions. Distinct topographic contrasts at opposite bias polarities, along with their spectroscopic characteristics, demonstrate that each defect species has a different chemical origin. At a low bias, the atomic resolution of graphene and the absence of defects confirm their subsurface nature [Supporting Information, S3].

Strikingly, at negative bias, the bright defects exhibit a wide range of dynamic behaviors from stable (indicated by a green circle) to unstable (red circle) configurations, as shown in Figure 2(a). They are of the same family, exhibiting similar shape and symmetry; however, the unstable species appears more frequently with scattered dots or lines when imaged at a negative bias below their occupied-state resonance bias (~-0.45 V). The instability is well reflected in the corresponding dI/dV map shown in Figure 2(b). Its intensity is most pronounced at bias voltages below -0.45 V, as evidenced by bias-dependent dI/dV maps in Figure 2(c).

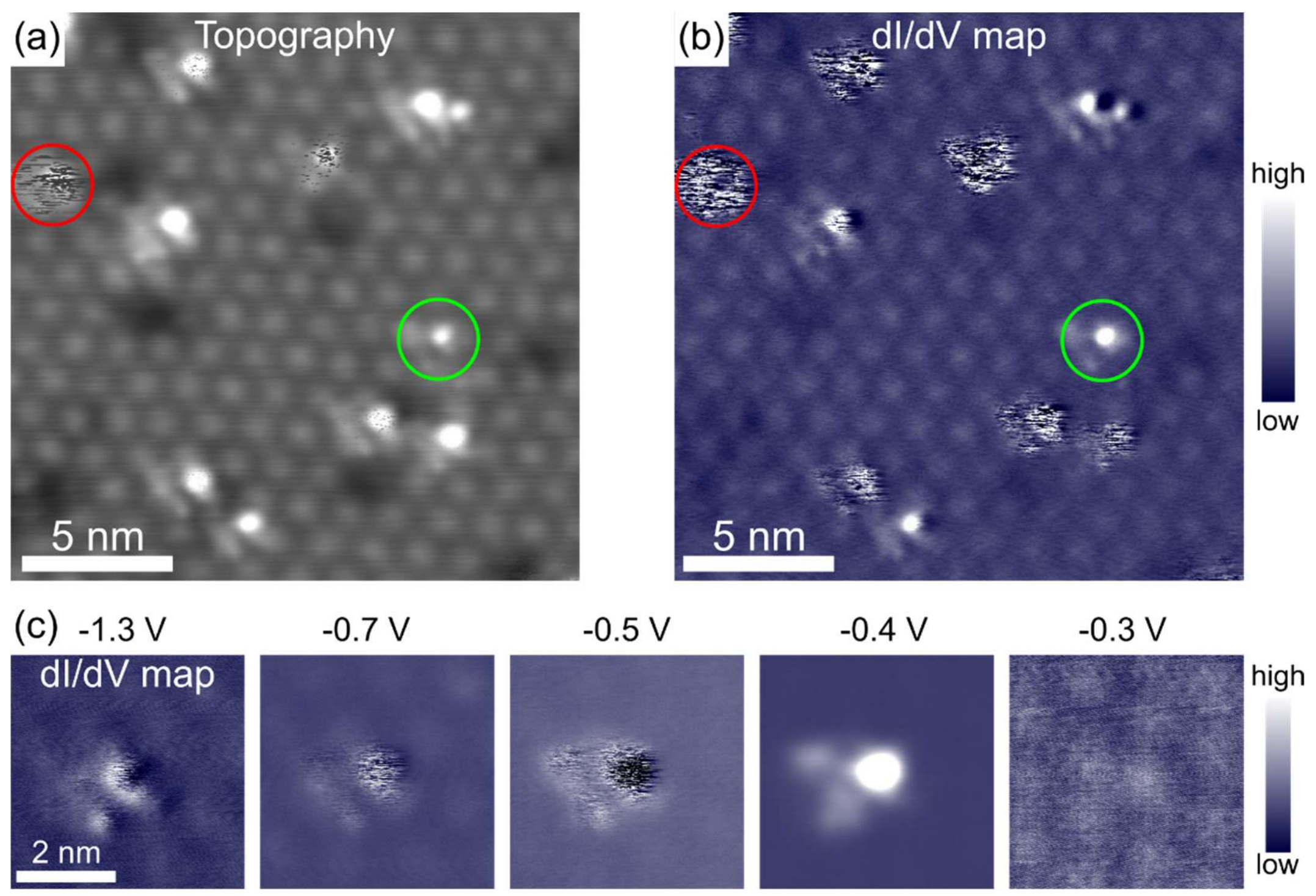

**Figure 2.** (a) Zoomed STM topography image (-1.3 V, 100 pA) of an intercalated Ag area. Four stable bright defects appear without noise (green circle), whereas the other seven defects exhibit instability with noise (red circle). However, the noise is only observed at a negative bias voltage below their resonance peak around -0.45 eV. (b) Corresponding spatial conductance dI/dV map taken at -1.3 V strongly reflects the instability of some bright defects in (a). (c) Bias-dependent dI/dV maps of a noisy defect at different negative biases; the noise is most intense at -0.5 V, which is below its resonance peak. The scale in the first image is applied to all the images.

Spectroscopy investigation further reveals the instability in bright defects. Figure 3(a) shows two dI/dV spectra acquired at a stable defect (navy) and an unstable defect (red). The instability at the unstable defect clearly emerges at negative bias below its resonance peak, consistent with the noise evolution in Figure 2(c). This instability is more evident in the corresponding I-V curve in Figure 3(b) where discrete low and high conductance levels can be seen (see zoomed inset). The histogram in the inset of Figure 3(a) reveals the bias threshold at which the current fluctuation starts, centered around -0.45 V, indicating that the tunnel junction polarity induces the instability. In contrast, spectra taken at the stable bright defect, dark defects, or a defect-free region show no instability.

To observe the tip-induced instability of bright defects in real time, the tip was positioned above a defect and a bias voltage at which switching occurs was applied between the tip and the sample with the feedback loop disabled, as illustrated in Figure 3(c). The corresponding tunneling current was recorded, enabling direct observation of defect excitation. Figure 3(d) represents a tunneling current trace recorded above an unstable defect at -0.5 V, identifying two conductance levels between low and high currents. Note that only two levels of conductance were observed on these defects, reflecting well-defined ON/OFF states, representing a bistable system identified as a nanoswitch[16,17,41]. We interpret this behavior as switching dynamics of the defects between two metastable configurations induced by tunneling electrons at negative bias even though the atomic configurations of these defects are unclear.

In strong contrast, no current fluctuations were observed at the stable bright defects or dark defects, which exhibit no noise in Figure 2, under positive biases using the same tip. This important observation indicates that the electron injection from the STM tip, assisted by a negative bias below the occupied-state resonance, facilitates switching in the bright defects at the measurement temperature of 5 K[42].

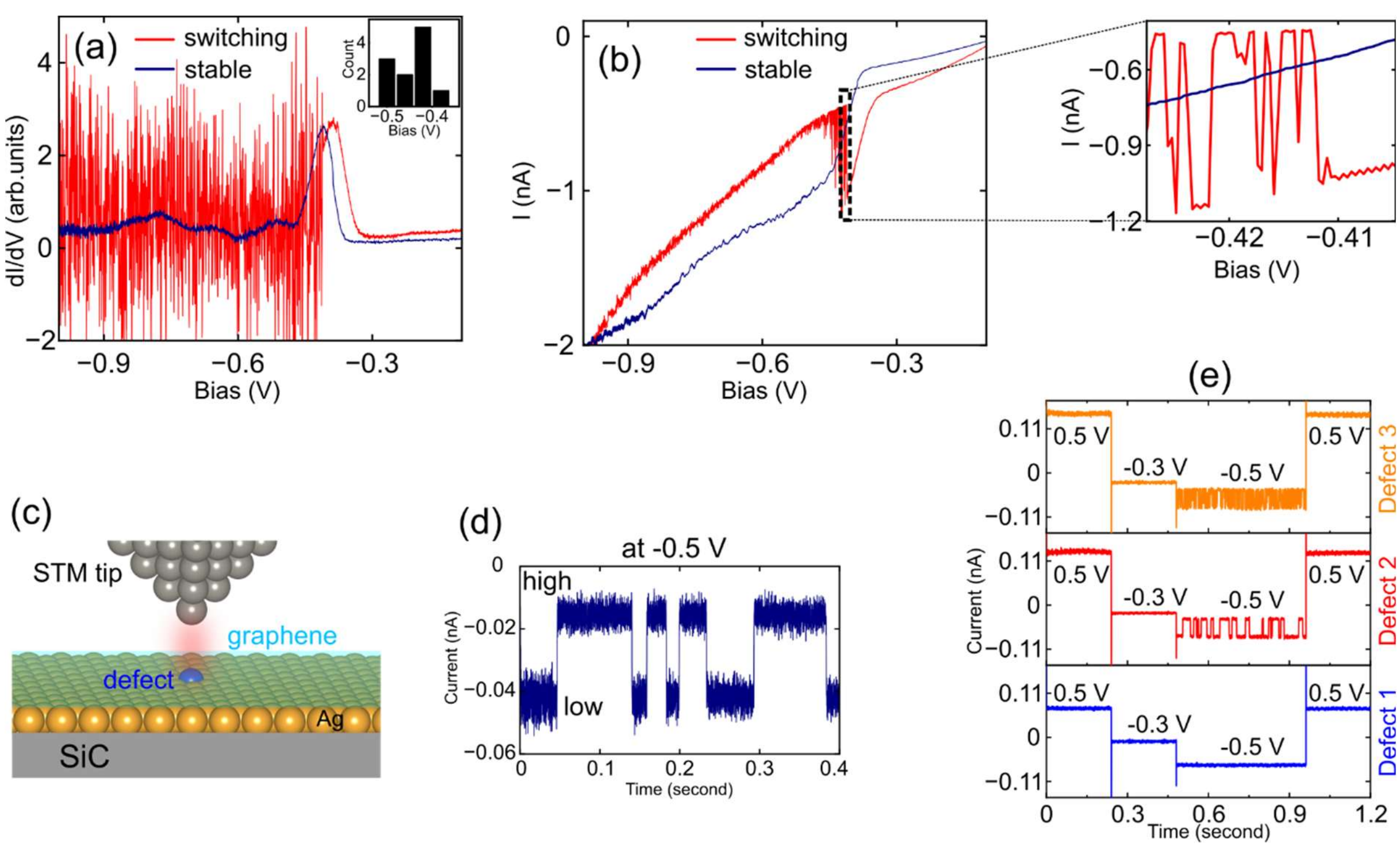

**Figure 3.** (a) dI/dV spectroscopy acquired above an unstable (red spectrum) and a stable bright defect (navy spectrum). Notably, a large noise is observed on the former defect, starting at a bias below its occupied state (for this specific defect is at -0.41 V). While the latter does not show instability with the same resonance peak, indicating that both defects are of the same type, however, adopting distinct dynamic behaviors. Inset: distribution of negative bias threshold at which the switching starts, observed over different measurements.

(b) Corresponding I-V trace in (a), revealing a strong current fluctuation in the bias range between -0.41 to -0.5 V (see zoomed region from the dashed rectangle). The large current instability with a smaller amplitude remains until the end of the spectrum toward the negative bias. (c) Schematic illustration of the tip-sample junction for controlling the switching of the defects in the Ag monolayer. The tip is placed above a defect with a fixed height; a negative bias (< -0.45 V) is applied and the current telegraph noise is recorded. The monolayer graphene is capped on top in light green. (d) Typical tunneling conductance telegraph trace recorded above a defect at -0.5V, characterized by a two-level conductance indicated as “low” and “high”. (e) Tunneling telegraph traces as a function of positive and negative bias voltages were recorded in three different bright defects, including one stable defect (blue curve, defect 1) and two other unstable defects (red curve for defect 2 and orange curve for defect 3). Along the spectra from left to right, a bias voltage of 0.5 V (tunneling current of 10 pA) is ramped, then switched to -0.3 V and -0.5 V stepwise, and again switched back to 0.5 V. Two-conductance tunneling level at -0.5 V indicates the switching observed only at this bias. In the case of a stable defect (blue spectrum), no switching events were observed through different biases. Note that the large step between different biases is due to the readjustment of the tip height in constant-current mode before recording the tunneling current with the disabled feedback loop.

To demonstrate various switching behaviors of bright defects, tunneling spectra were recorded on three individual bright defects (defects 1 to 3) using different bias voltages and polarities. The sample bias was sequentially varied from 0.5 V to -0,3 V, -0,5 V, and back to 0.5 V. Under identical tunneling conditions, defects 2 and 3 display distinct switching rates at -0,5 V under identical tunneling conditions.

To further elucidate the switching behavior of bright defects, we systematically investigated the dependence of the switching rate on the applied bias voltage, tunneling current, and tip-sample distance. First, we plot the switching rate as a function of the applied bias, shown in Figure 4(a). The rate (in Hz, where one period refers to the low-high-low switching cycle of the current) rises exponentially with the applied negative bias. Remarkably, no switching events were observed within the positive bias range, clearly indicating bias-polarity dependence. As previously shown, neutralization of charged atoms[42,43] can be driven through electron transfer induced by the STM tip. Similarly, the defect instability observed under negative bias in this work suggests that they are negatively charged. The applied negative bias at the tip-sample junction promotes defect neutralization, therefore, lowering the bonding barrier to the surface[42,44] and inducing switching. However, the defects are confined between the graphene and the Ag layer and cannot be removed by the tip.

The ability to control the defect switching process was further characterized as a function of the tunneling current. As shown in Figure 4(b), the switching rate rises exponentially, based on statistical measurements of four different defects at -0.5 V, each characterized by an exponential fit. These data were fitted to the power law $R \propto I^N$ where R is the switching rate (Hz), I is the applied tunnel current (pA), and N is the number of electrons. The exponential increase of the switching rate as a function of tunneling current provides strong evidence of an inelastic tunneling-induced tunneling process[16,18,45]. The fitted value N ranges from 1.83 to 2.44, indicating that the switching involves multiple electrons and varies among different defects, consistent with the different noise intensities seen in Figure 2(b). Such behavior is analogous to the number of electrons required to induce tautomerization in an organic molecule[17,18]. Consistently, Figure 4(c) plots the switching rate as a function of the tip-sample separation ($z$), where variations in $z$ modify both the tunneling current ($I \propto e^{-2\kappa z}$, $\kappa$ the tunneling decay constant) and the tip-induced electric field ($E \approx V/z$). As expected, the rate exhibits an exponential dependence on the tip height.

Notably, the switching rate varies drastically among defects because it depends exponentially on the negative bias [Figure 4(a)], while each defect has its own switching threshold determined by its occupied-state energy [inset of Figure 3(a)]. Consequently, a bias that activates one defect may induce much faster switching in others [Figure 3(e)]. These variations likely originate from differences in the local environment of defects, influenced by neighboring atoms, local strain, or inhomogeneous potential landscape imposed by the SiC substrate, as partly reflected in S3(c).

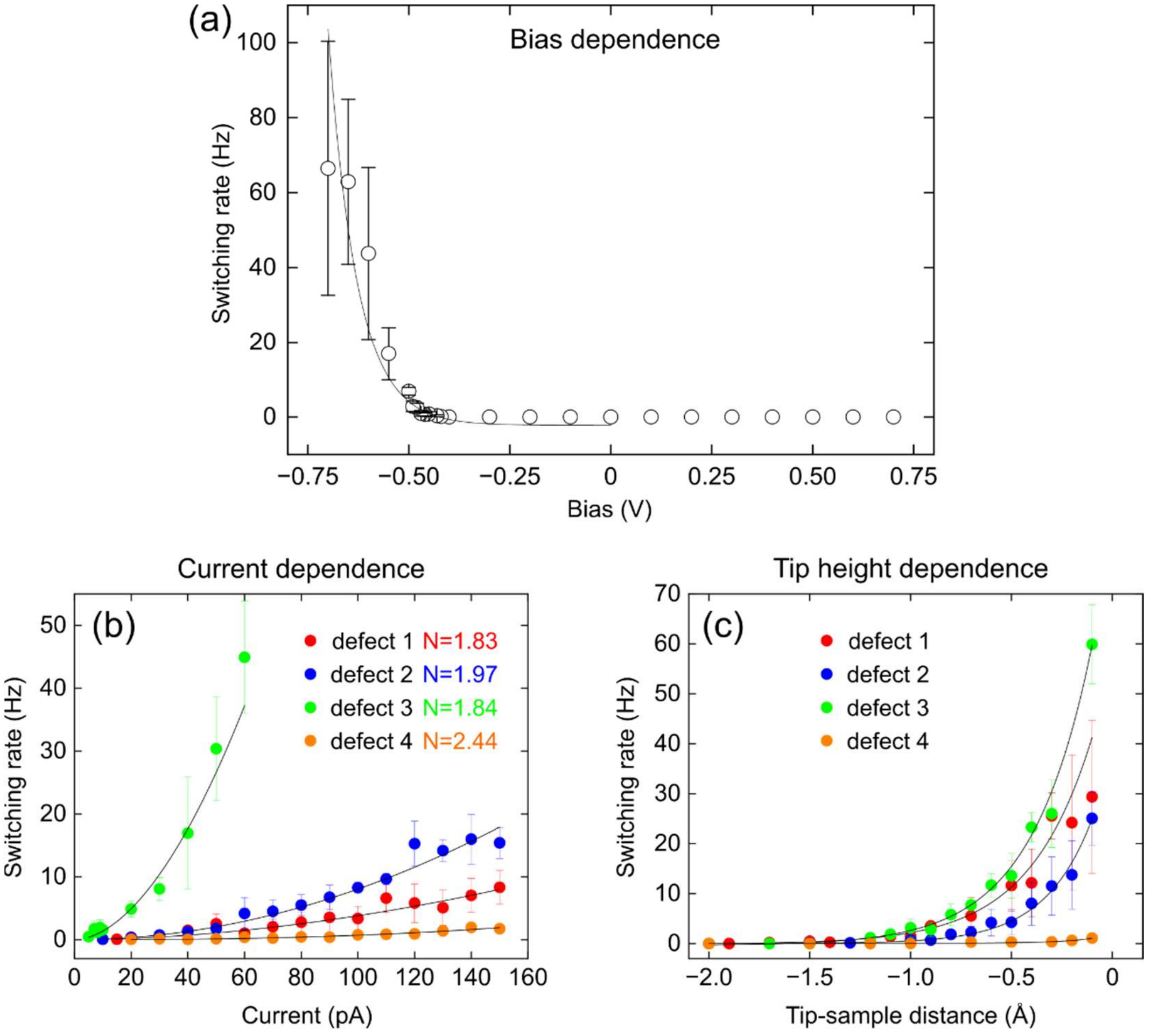

**Figure 4.** (a) Switching rate of bright defects measured as a function of bias voltage applied between the tip and the sample. Switching occurs only in the negative bias region, beginning just below the resonance state of the bright defects about -0.5 eV and the switching rate increases exponentially toward negative biases. The experimental points in the negative bias range are fitted by an exponential function (solid black curves). (b) Measured as a function of tunneling current taken on four different bright defects (initial tunneling conditions -0.5 V, 100 pA). The data are fitted to the exponential functions $R \propto I^N$ where R is the switching

rate (Hz), I is the applied tunnel current (pA), and N corresponds to the number of electrons participating in the switching process. We found that the number of electrons N varies between 1.83 to 2.44 for four example defects, clearly indicating an inelastic tunneling process. (c) Measured as a function of tip-sample separation acquired at -0.5 V. The experimental data points are fitted by an exponential fit (solid black curves).

These findings demonstrate that switching in bright defects can be effectively controlled by the negative bias, tunneling current, and tip-sample distance, enabling these defects to function as a nanometer-scale, electron-driven switching device[17]. Crucially, the atomically thin nature of graphene, combined with its weak interaction with Ag and the large momentum mismatch between tunneling electrons and the graphene band structure[46,47] acts as a nearly transparent tunneling barrier in the tip-defect junction. These exceptional properties of the graphene layer allow the STM tip to directly control the switching of the underlying defects in the Ag layer, while simultaneously confining the defects and preventing their removal by the tip. This creates a unique configuration for a defect-based nanoscale device at the graphene/SiC interface.

Switching leads to dramatic conformational change in bright defects, allowing one to gain insight into the switching process and explain the defect origins. Each bright defect exhibits its own switching rate. Others switch ultrafast, while some switch so slowly that their topographical modifications can be captured. Figure 5(a) shows four bright defects imaged at -0.73 V, in which two of them display fast switching, whereas the other two show a stable topography (left panel). In the consecutive STM image (right panel), the defect (marked by a black arrow) has switched and appears as a dark vacancy (Supporting Information, S5), clearly indicating that it is in the form of a substituted vacancy (e.g., Si[31,32,34]). This is evident in the corresponding constant-height current image as a sharp discontinuity in Figure 5(b), left panel. However, this switched configuration is temporary as the bright protrusion returns and reoccupies the exact site it initially vacated in the next scans (Supporting Information, S6). The above observation demonstrates that under tip excitation, the bright defect reversibly exchanges with the vacant site in the Ag layer.

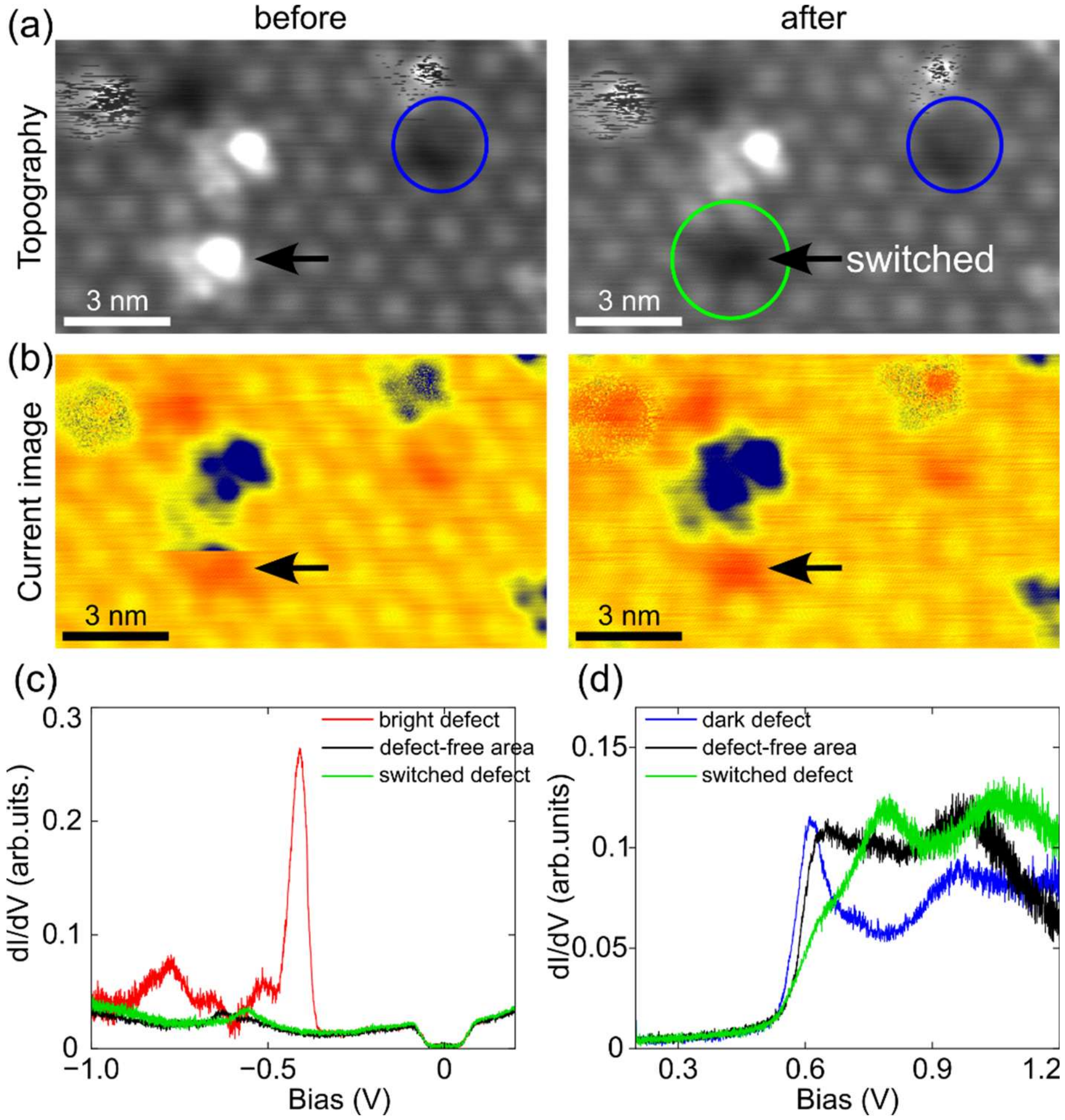

**Figure 5.** (a) STM topography image (-0.73 V, 100 pA) of an area with four bright defects. The blue circle indicates a nearby initial dark defect. After two consecutive scans, the lower bright defect (indicated by the black arrow) has transformed into a dark vacant site (indicated by the green circle, left panel); appearing similarly the initial dark defect (blue circle). The apparent heights of each defect species are indicated in Supporting Information, S5. More dynamic switching behaviors of these bright defects are detailed in Supporting Information, S6. (b) Constant-height current images (at -0.62 V) reveal the intermediate switching of the lower bright defect as a sharp discontinuity (black arrow, left panel). (c) dI/dV spectra in the negative bias range of the switched defect (green) as compared to those acquired above a stable defect

(red spectrum), defect-free area (black spectrum). The sharp resonance peak at -0.45 eV observed at the bright defect is absent in this switched configuration. (d) dI/dV spectra in the positive range taken at the switched defect are shown in the green spectrum. The spectra taken on an initial dark defect (blue spectrum), defect-free area (black spectrum) are plotted for comparison.

Spectroscopic investigations further support this conclusion. The green dI/dV spectrum taken at this switched defect is plotted in Figure 5(c,d) together with those taken on the nearby bright defect (red spectrum), the initial dark defect (blue spectrum), and the defect-free area (black) for comparison. At negative bias, the switched defect shows no electronic state near -0.45 eV, in contrast to that observed on the bright defects. At positive bias, it reveals features in the green curve in Figure 5(d)], which nearly resemble the electronic properties of the initial dark defect.

These findings confirm the presence of two distinct defect types in the monolayer Ag: the bright defects most likely arising from substitutional impurity at the Ag vacancies. The topographic and spectroscopic measurements clearly reveal that when a bright defect switches, it leaves behind an empty vacancy that is identical to the initial dark defect.

From these observations, we propose the following mechanism for the dynamic switching of bright defects. The bright defects, likely substitutional atoms occupying vacancies in the Ag layer, may carry a negative charge. A negative-bias tunneling current can transiently neutralize them, weakening their bonding with neighboring Ag atoms and enabling switching between two states (either vertically or laterally, although the exact pathway remains unclear), resulting in reversible two-state (ON/OFF) switching.

In conclusion, we employed scanning tunneling microscope to probe point defects in a silver monolayer intercalated at the graphene/SiC interface. Two distinct defect types were identified: bright and dark species; each associated with defect-induced states in the occupied and unoccupied energy ranges, respectively. The bright defects exhibit tip-induced switching dynamics, enabling atomic-scale two-level conductance controlled by tunneling current under negative bias, and serve as atomic-scale switches. Significantly, during switching, a bright defect transforms into a dark vacancy similar to the initial dark defect. We attribute the bright and dark defects to substitutional and empty vacancies, respectively, in the Ag monolayer (e.g., substitutional Si or other unknown impurities). This work highlights a first detailed insight into the dynamic behavior of defects in a two-dimensional intercalated metal probed by a scanning probe technique. It sheds light on defect identification and emerging defect-induced properties of intercalated silver at the graphene/SiC interface and establishes a foundation for the control of their properties with atomic precision.

## ASSOCIATED CONTENT

Additional analysis of the structures of monolayer Ag, provided in S1 and S2; supporting measurements of properties and dynamics of defects in monolayer Ag, detailed in S3, S4, S5, and S6.

## NOTES

The authors declare no competing financial interest.

## ACKNOWLEDGMENTS

The authors thank Audrey Gilbert (PDI) for careful reading and fruitful suggestions. J.A.R. and A.J. acknowledge the support of the National Science Foundation (NSF) Award No. DMR-2011839 (via the Penn State MRSEC-Center for Nanoscale Science). J.A.R and C.D. acknowledge the support by 2DCC-MIP under NSF cooperative agreement DMR-2039351.

# Supporting Information

**Table of Contents:**

## 1. Apparent height of monolayer Ag

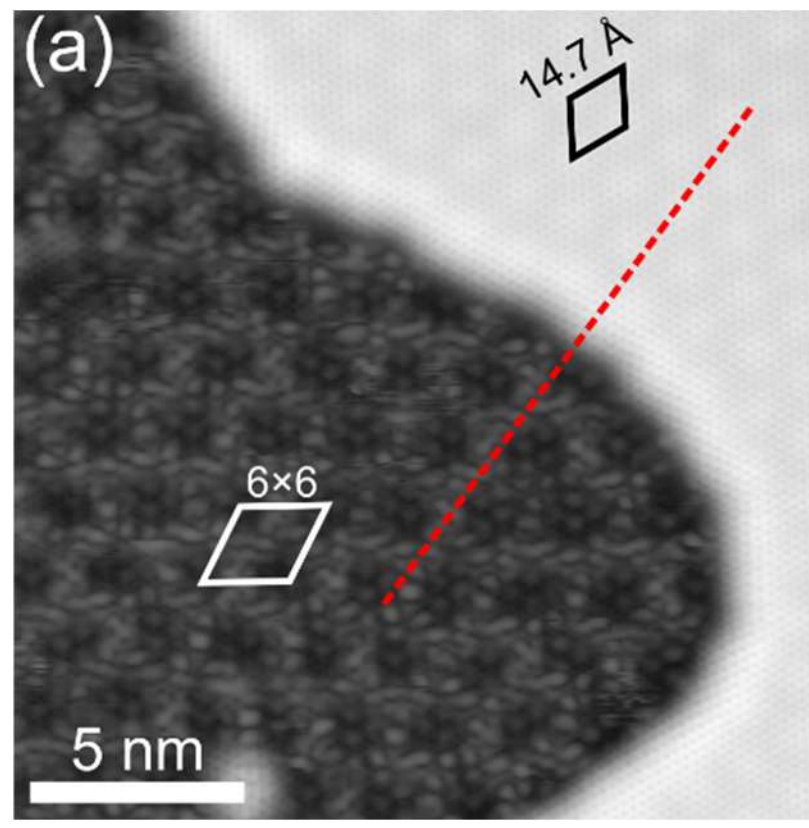

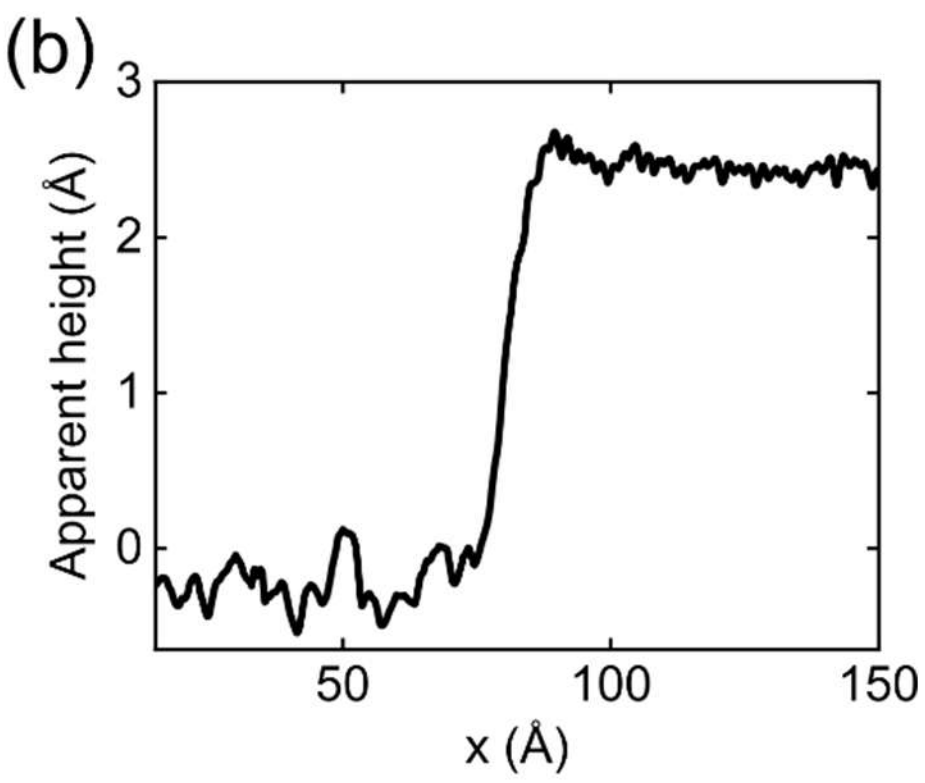

**S1.** (a) STM topography image of two regions between a pristine graphene/SiC (left side) and intercalated Ag (right side). The 6×6 superstructure (white rhombus) reveals a fingerprint that this area is composed of monolayer graphene on SiC without intercalants. The area on the right shows a periodicity of ~14.7 Å (black rhombus) as the result of the superposition of six graphene unit cells (2.46 Å) and five Ag unit cells (2.98 Å) in the Ag monolayer. (b) Line profile along the red dashed line in (a), revealing a height difference between them of ~3 Å, identifying an Ag monolayer intercalated below graphene.

Figure S1 shows a transition region between a pristine graphene/SiC and intercalated Ag. The 6×6 periodicity (white rhombus), which originates from the 6√3×6√3 reconstruction on SiC, indicates no Ag intercalated between the monolayer EG and SiC in this region. In contrast, the region on the right that exhibits a less corrugated surface with a periodicity of ~14.7 Å (black rhombus) is intercalated by Ag. The apparent height difference shown in S1 (b) between the two regions is measured of ~3 Å, revealing a monolayer Ag, which is in good agreement with previous findings[1].

## 2. Moiré pattern formed between 2D Ag and graphene

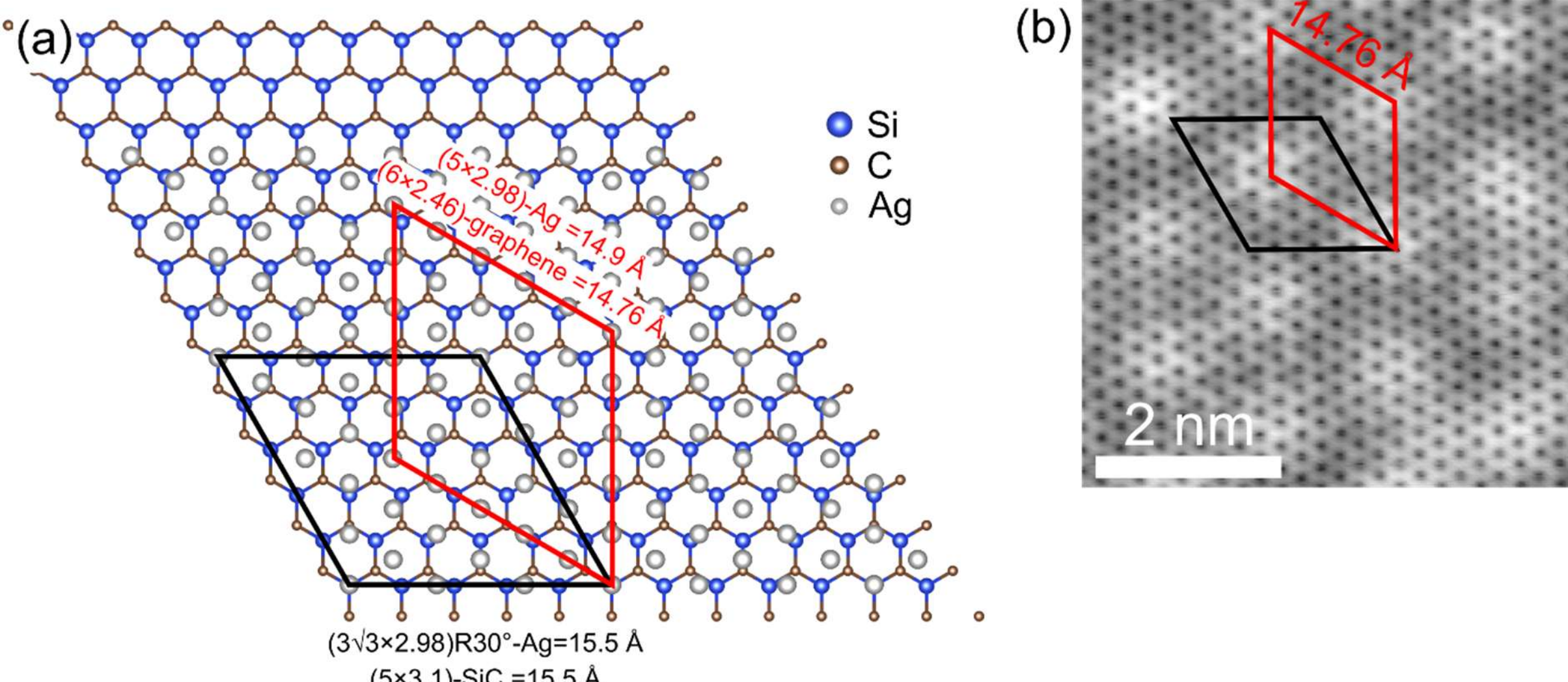

**S2.** (a) (3√3×3√3)R30°-Ag supercell fitting a (5×5)-SiC supercell indicated by the black rhombus[2]. The graphene layer, which is aligned with the Ag lattice, is not included for a better illustration. The moiré pattern (lattice constant of 14.7 Å, indicated by the red rhombus) observed in the experiment shown in (b), results from the superposition of six graphene unit cells (2.46 Å) and five

Ag unit cells (2.98 Å) in the Ag monolayer. Both Ag and graphene are aligned. The black rhombus supercell is not detectable by STM imaging. (b) Experimental STM topography image (0.05 V, 100 pA) of intercalated Ag exhibiting a moiré pattern with a periodicity representing the red rhombus shown in (a).

Fig. S2(a) represents the (3√3×3√3)R30°-Ag supercell fitting a (5×5)-SiC supercell (black rhombus). The moiré pattern, which is shown in the STM image of Fig. S2(b) [which is also shown in Fig. 1 (a) and 2(a)], results from the superposition of six graphene unit cells and five Ag unit cells (2.98 Å) of the underlying monolayer, characterized by the 14.7 Å (depicted by red rhombus in Fig. S2 (a). In Fig. S2(a), the graphene overlayer is not displayed for better clarity). This moiré pattern periodicity represents the size of the red rhombus shown in (a). Note that the graphene remains rotated by 30° with respect to the SiC lattice after intercalation. This is because the intercalation only lifts the graphene buffer layer without inducing an additional rotation, consistent with LEED measurements reported in previous studies[3,4]. From this moiré pattern periodicity (14.7 Å), we deduced an epitaxial matching between Ag and SiC, which leads to the conclusion that each (3√3 × 3√3)-Ag supercell fits well with a (5 × 5)-SiC supercell with an Ag-Ag spacing of 2.98 Å, depicted by the black rhombus in Fig. S2 (a). Note that the black supercell was not detectable in the STM image in Fig. S2 (b).

The lattice constant of 2.98 Å in the monolayer Ag was further confirmed by LEED measurements and DFT calculations as it exhibits a most energetically favorable configuration[2]. We define this epitaxial matching (27:25)-Ag structure according to the Ag:SiC matching ratio and mentioned in the main text.

### 3. Bias-dependent imaging of defects in 2D Ag

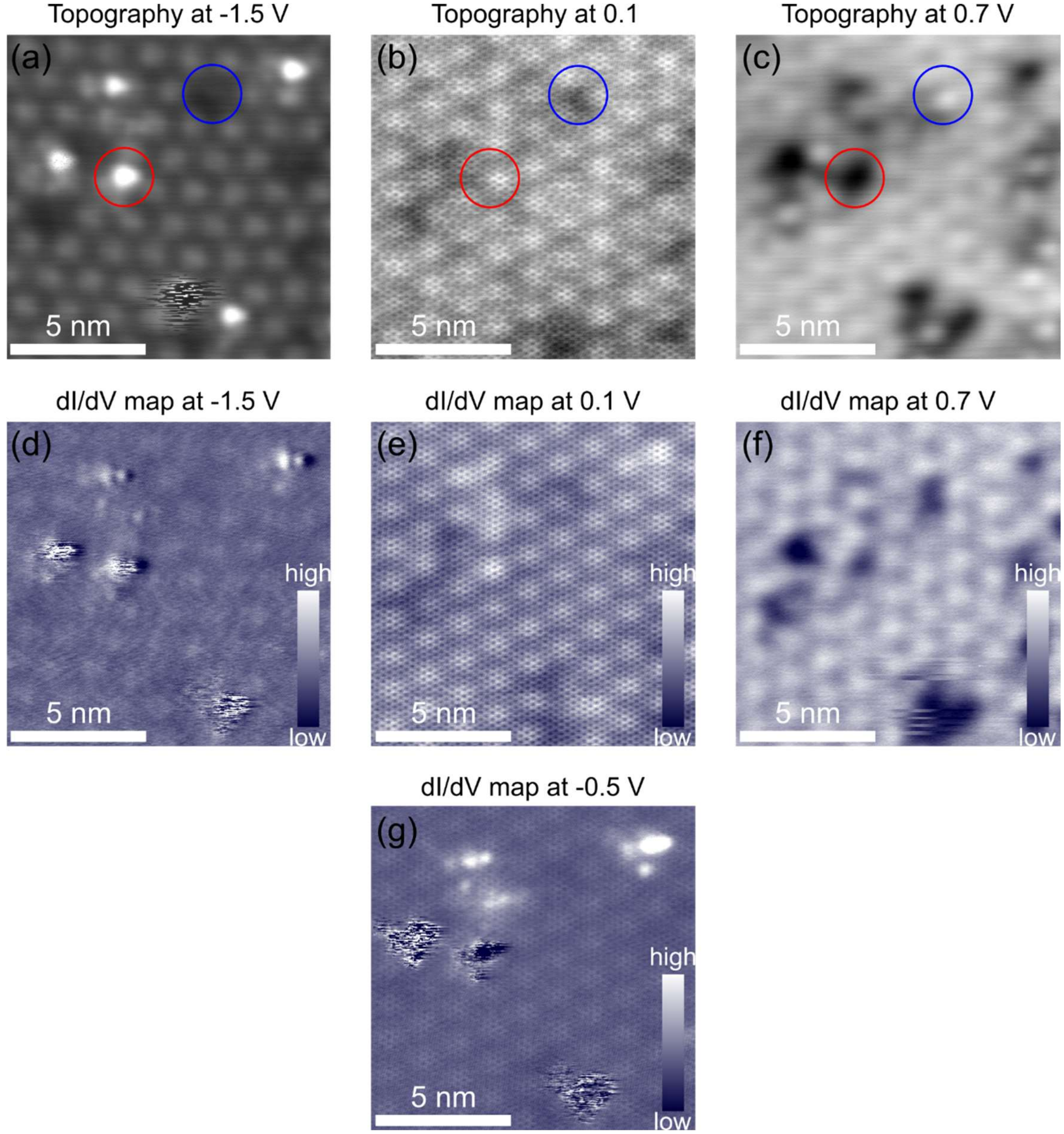

**S3.** (a-c) STM topography images of an intercalated Ag area taken at different bias voltages, revealing bias-dependent contrast of bright and dark defects, indicated by red and blue circles, respectively. (d-f) Corresponding dI/dV maps acquired with a constant-tip height at the same area as in (a-c), respectively. At 0.1 V, the area shows only a graphene honeycomb lattice. This indicates that the defects are located well below the graphene layer and not adsorbed on top of it. The bias-dependent topography contrast of the defects and their dI/dV density of state maps reflects bias-dependent wavefunction overlaps of intercalated Ag and graphene with the STM tip. Note that the graphene is also visible at -0.5 V at which the defects are clearly observed in S3 (g).

The satellite spots adjacent to the bright defects, observed at -1.5 V and -0.5 V, vary from defect to defect in both brightness and number. We tentatively attribute these features to possible charge transfer from the defect to the nearby Ag atoms. Depending on the spatial extent of this charge redistribution, which may vary with local environment, one, two or several Ag atoms could be affected and become visible as small bright spots.

## 4. Tip height dependence of dI/dV over the bright defects

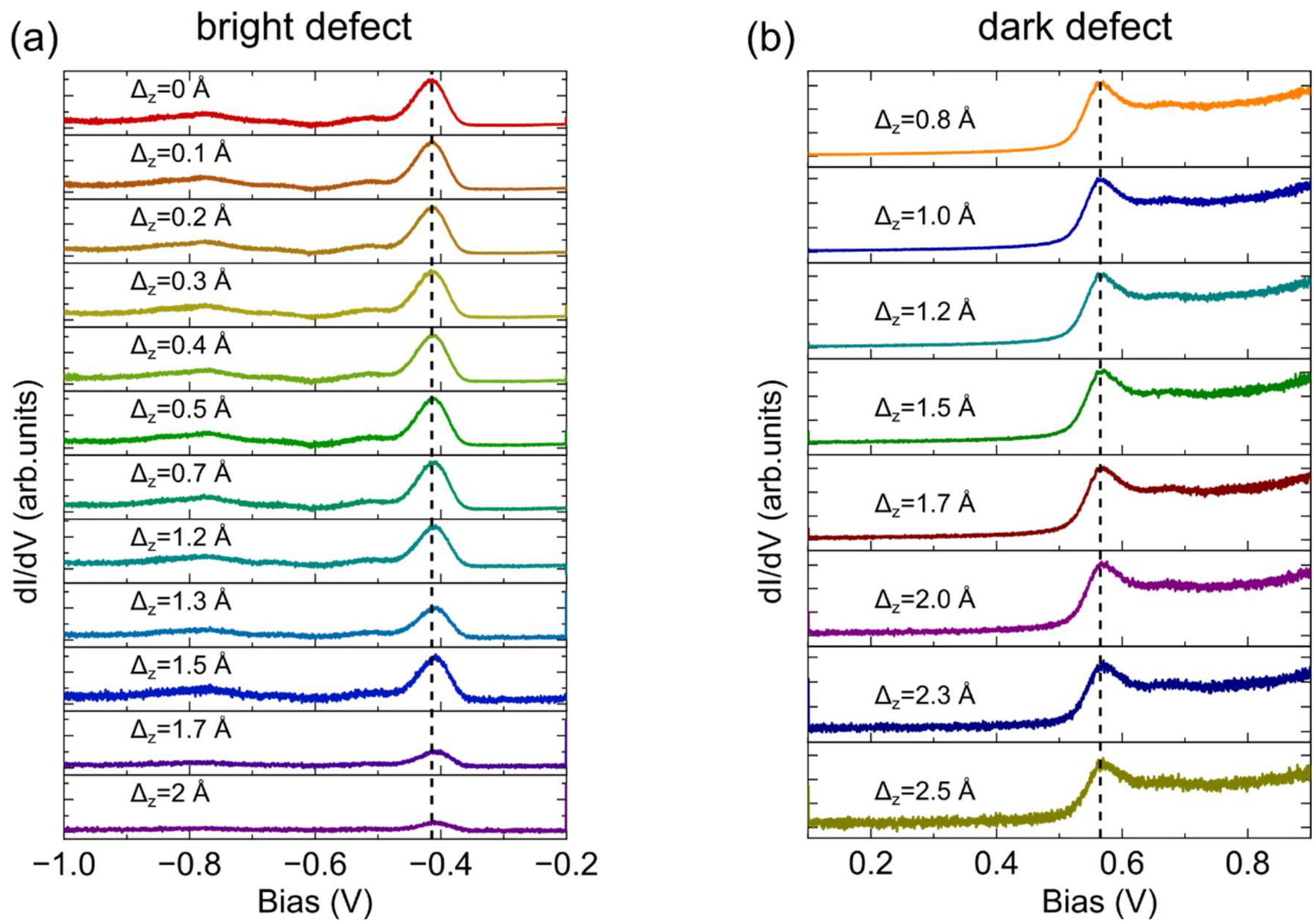

**S4.** Series of dI/dV spectra acquired on a bright defect (a) and on a dark defect (b) with varying tip-sample variation ($\Delta_z$) (indicated above each spectrum at an initial tunneling condition: 0.1 V, 100 pA). Note that tip-sample variation ($\Delta_z$) is relative to the initial tip height, defined by the set point. $\Delta_z$=0 Å implies that the spectra were acquired without changing a z-offset and $\Delta_z$=0.1 Å, for instance, implies that the tip is retracted 0.1 Å away from the surface. In both cases, the occupied and unoccupied peaks show almost no shift as a function of tip-sample distance. A small progressive shift over the range of ~2 Å is observed, but this is not an indication of the charging peak.

## 5. Height profile of each defect species

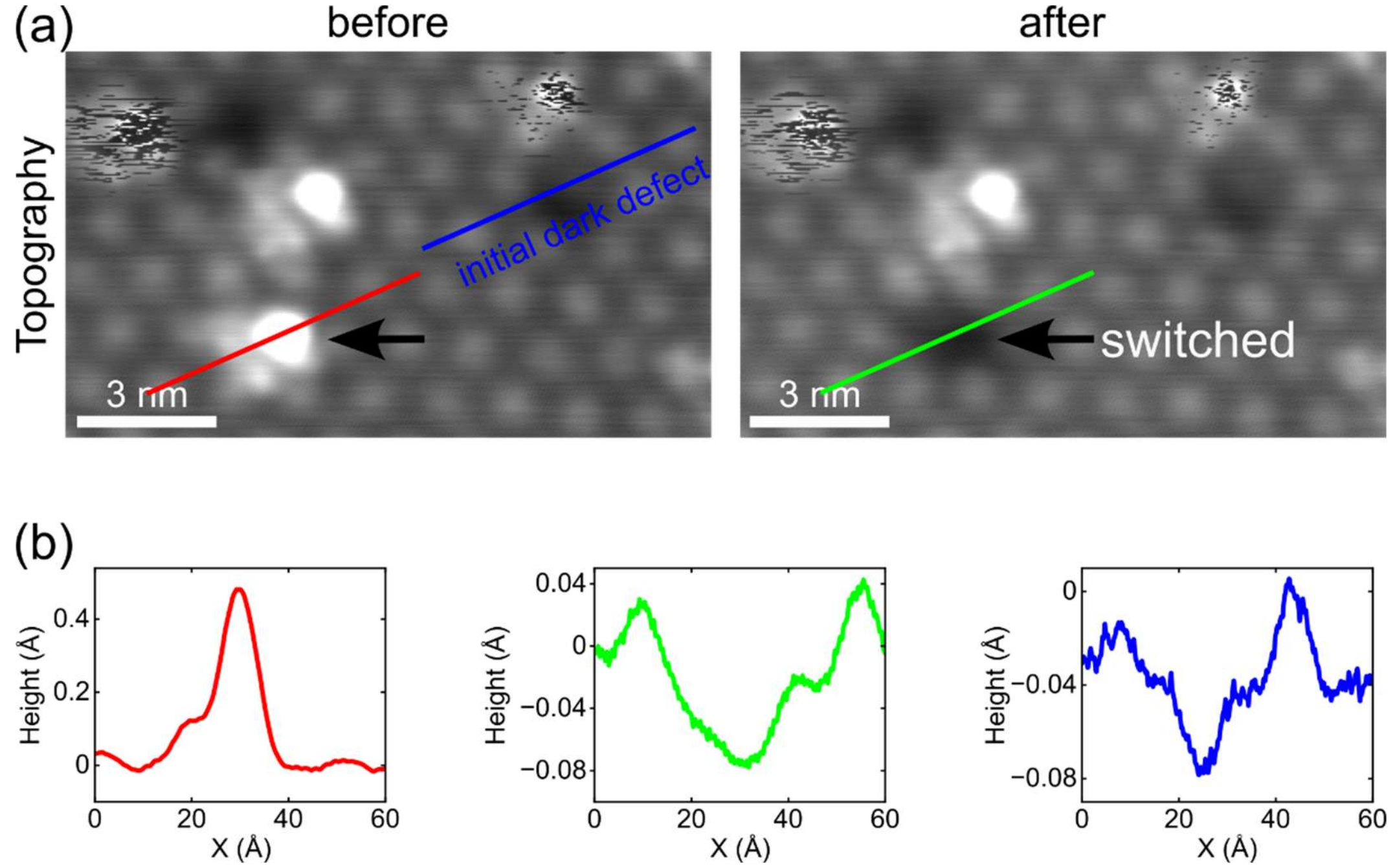

**S5.** (a) STM images (-0.73 V, 100 pA) of bright and dark defects before (left panel) and after (right panel) the lower bright defect switched (indicated by the black arrow). (b) Corresponding height profiles of the lower bright defect before switching (red curve) and after switching (green curve) in comparison with the height of an initial nearby dark defect (blue curve).

## 6. Consecutive STM images of switching dynamics in bright defects

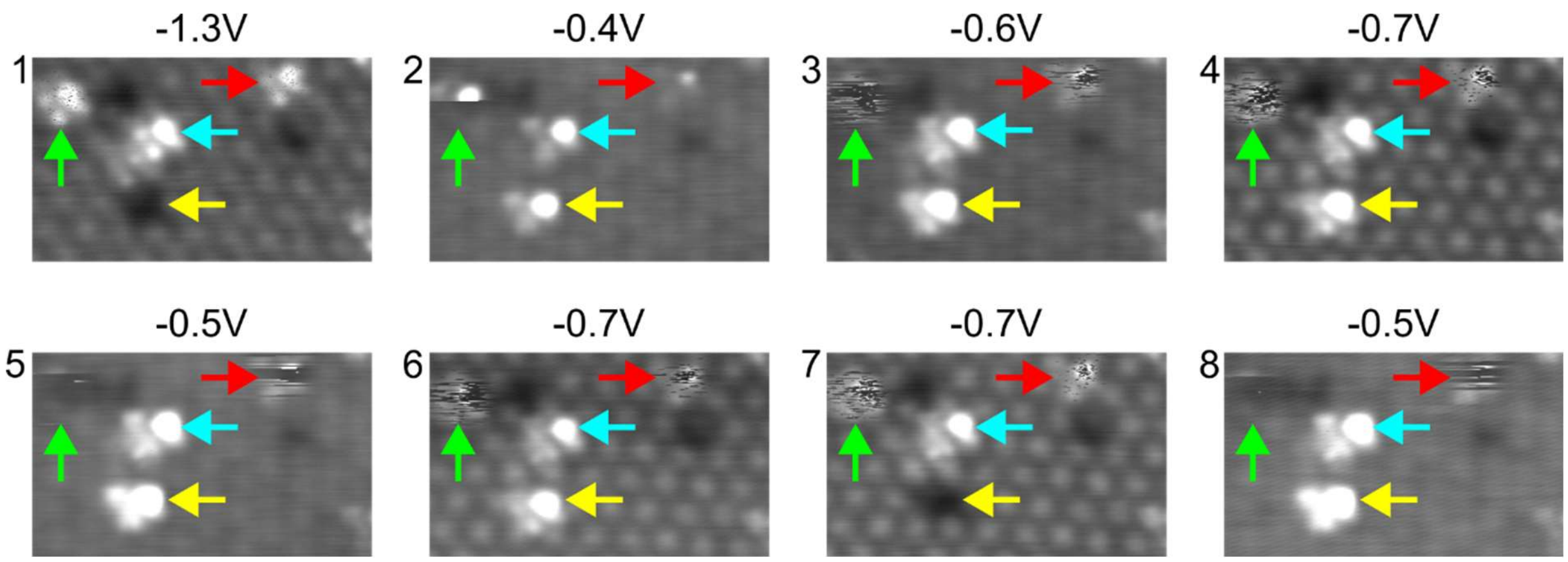

**S6.** Sequence of STM images (labelled 1-8) acquired at different biases over time, showing different switching behaviors of four bright defects indicated by green, red, cyan, and yellow arrows under the same tip condition. The defect marked by the cyan arrow remains stable across the image sequence, whereas three other bright defects (indicated by green, red, and yellow arrows) exhibit fast and slow switching rates. Upon switching, the defects marked by green and yellow arrows transform into dark vacant sites.

## References for Supporting Information